\documentclass[conference]{IEEEtran}
\IEEEoverridecommandlockouts

\usepackage{subfigure}

\usepackage{graphicx}

\ifCLASSINFOpdf
   \usepackage[pdftex]{graphicx}
\else
\fi
\usepackage[cmex10]{amsmath}
\usepackage{amssymb}  
\usepackage{amsfonts}

\usepackage{array}
\begin{document}
%
\title{LQG feedback control of a class of linear non-Markovian quantum systems}

%

\author{Shibei~Xue, Matthew~R.~James, Valery Ugrinovskii, and~Ian~R.~Petersen
\thanks{This research was supported under Australian Research Council¡¯s Discovery
Projects and Laureate Fellowships funding schemes (Projects DP140101779 and FL110100020) and the Chinese Academy of Sciences President's International Fellowship Initiative (No. 2015DT006).}
\thanks{S. Xue, V. Ugrinovskii and I. R. Petersen are with the School of Information Technology and Electrical Engineering, University of New South Wales Canberra at the Australian Defence Force Academy, Canberra, ACT 2600, Australia (e-mail: xueshibei@gmail.com; v.ugrinovskii@gmail.com; i.r.petersen@gmail.com).}
\thanks{M. R. James is with the ARC Centre for Quantum Computation and
Communication Technology, Research School of Engineering, Australian
National University, Canberra, ACT 0200, Australia (e-mail: Matthew.James@anu.edu.au).}
}

%


\maketitle

\begin{abstract}
In this paper we present a linear quadratic Gaussian (LQG) feedback control strategy for a class of linear non-Markovian quantum systems. The feedback control law is designed based on the estimated states of a whitening quantum filter for an augmented Markovian model of the non-Markovian open quantum systems. In this augmented Markovian model, an ancillary system plays the role of internal modes of the environment converting white noise into Lorentzian noise and a principal system obeys non-Markovian dynamics due to the direct interaction with the ancillary system. The simulation results show the LQG controller with the whitening filter obtains a better control performance than that with a Markovian filter in the problem of minimizing the photon numbers of the principal system when the ancillary system is disturbed by thermal noise.
\end{abstract}


%
\IEEEpeerreviewmaketitle

\section{Introduction}
%
%
%
%
%

Feedback is a fundamental technique in classical control engineering and has been applying to quantum systems as well, for example, stabilizing a quantum system in an eigenstate by using Lyapunov feedback control~\cite{Mirrahimi} and rejecting noise in a linear quantum system by using $H^\infty$ feedback control~\cite{hinfinity}, etc.

The feedback controlled quantum systems in most existing works are Markovian quantum systems. They refer to quantum systems disturbed by  memoryless environments~\cite{bouten,Wiseman1994,HarocheNAT2011,Altafini2012,Breuer,Gardiner}, where the correlation time of the environment is much larger than that of the quantum system.  Thus the noise from the environment can be assumed to be white noise satisfying a delta-commutative relation~\cite{Gardiner}. Here, the filter used for the feedback control is designed based on the Markovian system model. However, when a complicated environment, e.g., a non-Markovian environment involving colored noise, is considered, the existing feedback strategies will induce degraded performance. The main reason for this is that the quantum filters or observers used for feedback control of Markovian quantum systems have not taken the internal dynamics of the non-Markovian environment into account.

In classical control engineering, we can always append the state of the colored noise to the state of the system so as to obtain an augmented model of the total system which is only driven by white noise~\cite{Hanggi,KS72}. Based on this augmented model, a whitening filter can be constructed via observing  the output of the system. For example, considering an augmented classical dynamical system
\begin{eqnarray}\label{000}
  \dot x_1(t)&=&-x_1(t)+x_2(t),\nonumber\\
  \dot x_2(t)&=&-x_2(t)+\nu(t),
\end{eqnarray}
the evolution of the system state $\left[
                                     \begin{array}{cc}
                                       x_1(t) & x_2(t) \\
                                     \end{array}
                                   \right]^T$ is apparently Markovian driven by white noise $\nu(t)$, i.e., the variation of the state is determined by the current state. While the dynamics of the subsystem $x_1(t)$ is non-Markovian due to the injected colored noise from the subsystem $x_2(t)$. Similar ideas also arise in quantum systems, e.g., the pseudo-mode method~\cite{PhysRevA.50.3650,PhysRevA.80.012104} and the hierarchy equation approach~\cite{PhysRevA.85.062323,shabani2014} for modelling non-Markovian quantum systems, and the estimation of the quantum system driven by non-classical fields~\cite{gough2014NJP,gough2013QIP,gough2012PRA}. For effectively estimating the states of linear non-Markovian quantum systems, a whitening quantum filter is presented in~\cite{xue2015} based on an augmented system model. An ancillary system defined on a Hilbert space $\mathcal{H}_a$ is introduced, which takes the effects of the internal modes of the non-Markovian environment converting white noise into colored noise, and it is directly coupled with a principal system defined on a Hilbert space $\mathcal{H}_p$. Hence, the total system evolves on the Hilbert space $\mathcal{H}_p\otimes\mathcal{H}_a\otimes\mathcal{F}$, where $\mathcal{F}$ is a Fock space for white noise. A standard \textsl{Belavkin} quantum filter can be obtained for this augmented system by applying a non-demolition probing field to the principal system. Thus the estimated states of the principal system can be used in the feedback control of the principal system.

In this paper, we utilize the estimated states of the whitening quantum filter in the feedback control of a linear and Lorentzian-noise-disturbed non-Markovian quantum system. Since both the Lorentzian noise generator; i.e., the ancillary system, and the principal system are linear quantum systems, the feedback control law can be designed for the augmented system by using a linear quadratic Gaussian (LQG) approach. The optimal controller includes the whitening quantum filter for reconstructing the states of the non-Markovian quantum system and a linear function of the estimated states~\cite{KS72}. The separation principal guarantees that the quantum filter and the feedback control law can be designed separately~\cite{KS72}.

The paper is organized as follows. Markovian quantum systems are briefly reviewed in Section~\ref{sec2}. In Section~\ref{sec3}, an augmented system model for a Lorentzian-noise-disturbed non-Markovian quantum system is introduced. Based on the augmented system model, a whitening quantum filter for the non-Markovian quantum system is given in Section~\ref{sec4}. The LQG feedback control for this linear non-Markovian quantum system is presented in Section~\ref{sec5}. A possible experimental realization using optical systems and corresponding simulation results are shown in Section~\ref{sec6}. Conclusions are drawn in Section~\ref{sec7}.
\section{Review of Markovian quantum systems}\label{sec2}
\subsection{$(S,L,H)$ description}
A Markovian quantum system $G$ refers to a quantum system interacting with white noise fields,
which can be systematically described by an $(S,L,H)$ description as
\begin{equation}\label{13-11}
  G=(S,L,H),
\end{equation}
where the component $S$ is a scattering matrix describing the input-output relation of fields passing through beam splitters, the operator vector $L$ is a collection of system operators interacting with the white noise fields, and $H$ is the system Hamiltonian~\cite{Gough2009}.
\subsection{White noise field}
The white noise field on the Boson Fock space $\mathcal{F}$ can be defined as
\begin{equation}\label{5}
  b(t)=\frac{1}{\sqrt{2\pi}}\int_{-\infty}^{+\infty}b(\omega)e^{-{\rm i}\omega t}{\rm d}\omega
\end{equation}
satisfying the delta commutation relations
\begin{equation}\label{3}
[ b(t), b^\dagger(t')]=\delta(t-t'), [ b(t), b(t')]=0.
\end{equation}

From the definition (\ref{5}), an integrated operator can be defined as $B_t=\int_{t_0}^t b(t'){\rm d}t'$, $B_t^\dagger=\int_{t_0}^t b^\dagger(t'){\rm d}t'$
whose commutation relation can be calculated as $[B_t,B_{t'}^\dagger]={\rm min}(t,t'), [B_t,B_{t'}]=0$ and thus the operator $\Theta_t=B_t+B_t^\dagger$ is the quantum analog of the Wiener process and $\theta(t)=b(t)+b^\dagger(t)$ is quantum white noise. Also a scattering process can be defined as $\Lambda_t=\int_{t_0}^tb^\dagger(t')b(t'){\rm d}t'$. Note that we assume the initial state of the field on the Fock space $\mathcal{F}$ is a vacuum state such that this process is analogous to Gaussian white noise with zero mean.
\subsection{Quantum stochastic differential equation}
The evolution of the system $G$ satisfies a quantum stochastic differential equation (QSDE)
\begin{eqnarray}\label{10}
  {\rm d}U_t&=&{\big\{}-{\big(}{\rm i}H+\frac{1}{2}L^\dagger L{\big)}{\rm d}t+{\rm d}B_t^\dagger L\nonumber\\
  &&~~~~~~~~~~~~~~~-L^\dagger S{\rm d}B_t+(S-I){\rm d}\Lambda_t{\big\}}U_t.
\end{eqnarray}



In the Heisenberg picture, the evolution of a system operator $X$ can be defined as $j_t(X)=U^\dagger(t)XU(t)$ satisfying a QSDE , namely a quantum Langevin equation, as
\begin{eqnarray}\label{11}
 {\rm d}j_t(X)&=&j_t(\mathcal{G}(X)){\rm d}t+{\rm d}B_t^\dagger j_t(S^\dagger[X,L])\nonumber\\
 &&+j_t([L^\dagger,X]S){\rm d}B_t+j_t(S^\dagger XS-X){\rm d}\Lambda_t,
\end{eqnarray}
with a generator
\begin{equation}\label{11-1}
\mathcal{G}(X)=-{\rm i}[X,H]+\mathcal{L}_L(X).
\end{equation}
The notation $\mathcal{L}_{\cdot}(\cdot)$ defines a Lindblad superoperator which can be calculated as $\mathcal{L}_{R}(O)=\frac{1}{2}R^\dagger[O,R]+\frac{1}{2}[R^\dagger,O]R$ for two arbitrary operators $R$ and $O$ with suitable dimensions.
Such an equation describes the dynamics of the system driven by an external white noise field, which has been widely used in the analysis and control of Markovian quantum systems~\cite{Gardiner}.


\subsection{Input-output relations}\label{inputoutput}

To observe the dynamics of the system, one may consider an output field. The output field is the field after interaction with the system, which satisfies a QSDE as
\begin{equation}\label{13}
  {\rm d}B_{\rm out}(t)=j_t(L){\rm d}t+j_t(S){\rm d}B_t,
\end{equation}
which shows that the output field not only carries information of the system but also is affected by the noise. As a result, the output field can be utilized by a quantum filter or a feedback controller~\cite{bouten,MY2003,ZGF2011}.

\section{Markovian representation for a class of non-Markovian quantum systems disturbed by Lorentzian noise}\label{sec3}
In this section, we consider a class of non-Markovian quantum systems disturbed by Lorentzian noise. Instead of describing the system in a traditional way~\cite{Tu2008,xue2013}, we will introduce an augmented system model. In this model, colored noise in the non-Markovian environment is generated by an ancillary system driven by white noise and a principal system is directly coupled with the  ancillary system resulting in its non-Markovian dynamics. The structure of the augmented system model is given in Fig.~\ref{AP}.
\subsection{Ancillary system driven by white noise}
To convert white noise to colored noise with a Lorentzian spectrum, we consider an ancillary system is described by
\begin{equation}\label{}
  G_a=({\rm I},\sqrt{\gamma_0}a_0, \omega_0a^\dagger_0 a_0);
\end{equation}
i.e., an optical mode in a leakage cavity, where $\omega_0$ is the angular frequency and $a_0$ ($a_0^\dagger$) is the annihilation (creation) operator of the ancillary system defined on a Hilbert space $\mathcal{H}_a$. Here the coupling operator is chosen as $\sqrt{\gamma_0}a_0$, where $\sqrt{\gamma_0}$ is a damping rate with respect to the white noise field.

With respect to $G_a$, a QSDE for the annihilation operator $a_0$ can be obtained as
\begin{equation}\label{14}
 {\rm d}a_0(t)=-(\frac{\gamma_0}{2}+{\rm i}\omega_0)a_0(t){\rm d}t-\sqrt{\gamma_0}{\rm d}B_t.
\end{equation}
We define
\begin{equation}\label{14-1}
 c(t)=-\frac{\sqrt{\gamma_0}}{2}a_0(t)
\end{equation}
as a fictitious output.

It is easy to check that
the power spectral density for  $c(t)$ is Lorentzian calculated to be
\begin{equation}\label{19}
  S(\omega)=\frac{\frac{\gamma_0^2}{4}}{\frac{\gamma_0^2}{4}+(\omega-\omega_0)^2}
\end{equation}
with a center frequency $\omega_0$ and a linewidth $\gamma_0$ determined by the angular frequency of the ancillary system and the damping rate with respect to the white noise field, respectively.
%
%
Note that
in the broadband limit~\cite{Gough06JMP}; i.e., $\gamma_0\rightarrow\infty$,
the fictitious output $c(t)$ reduces to white noise with a delta correlation function~\cite{xue2015}.

\subsection{Principal system interacting with the ancillary system}
A principal system $G_p$ on a Hilbert space $\mathcal{H}_p$ with a free Hamiltonian $H_P$ is of interest, which is disturbed by the colored noise created by the noise model, i.e., the ancillary system, via a direct interaction. Thus the principal system and the ancillary system constitute an augmented system.

\begin{figure}
  \includegraphics[width=8.5cm]{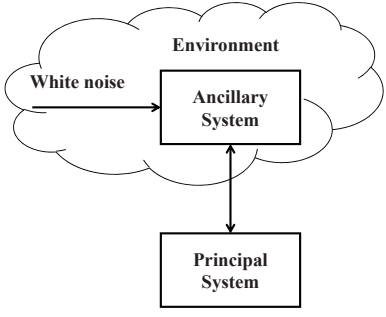}\\
  \caption{Schematic diagram for the direct coupling between the ancillary and the principal system~\cite{xue2015}.}\label{AP}
\end{figure}

The interaction Hamiltonian for the coupling between the principal system and the ancillary system is
\begin{equation}\label{22}
  H_I={\rm i}(c^\dagger Z-Z^\dagger c),~~~~Z=\sqrt{\kappa}K,
\end{equation}
where $Z$ is a direct coupling operator of the principal system expressed as
a product between a principal system operator $K$ and a coupling strength $\sqrt{\kappa}$. The principal and ancillary systems influence each other due to their direct interaction as shown in Fig.~\ref{AP}.
This augmented principal-ancillary system can be described by using an $(S,L,H)$ description as
\begin{equation}\label{23}
 G_{p,a}=({\rm I},\sqrt{\gamma_0}a_0,H_P+H_I+\omega_0a^\dagger_0a_0).
\end{equation}

By cancelling the operators of the ancillary system, a non-Markovian Langevin equation for an operator $X$ of the principal system can be obtained as
\begin{eqnarray}\label{32-1}
   \dot{\bar{X}}_t &=& -{\rm i}[\bar{X}_t,\bar{H}_S(t)]+c^\dagger(t)[\bar{X}_t,\bar{Z}_t]+[\bar{Z}_t^\dagger,\bar{X}_t]c(t)\nonumber\\
   &&+D(\xi^*,\bar{Z}^\dagger)_t[\bar{X}_t,\bar{Z}_t]+[\bar{Z}_t ^\dagger,\bar{X}_t]D(\xi,\bar{Z})_t
\end{eqnarray}
where the convolution terms are expressed as
\begin{eqnarray}\label{32-2}
  D(\xi,\bar{Z})_t&=&\frac{1}{2}\int_{t_0}^t \xi(t-\tau)\bar{Z}_\tau{\rm d}\tau.
\end{eqnarray}
Note that we label corresponding operators with a bar in this augmented system picture.

This Langevin equation coincides with the existing non-Markovian Langevin equations whose integral terms represent the memory effect~\cite{XuePRA2012,Tan2011}. Note that in the broadband limit $\gamma_0\rightarrow+\infty$, (\ref{32-1}) reduces to a standard Markovian Langevin equation~\cite{Gardiner}.

\section{Whitening quantum filtering for non-Markovian quantum systems}\label{sec4}


To estimate the dynamics of the non-Markovian system, we can construct a quantum filter using a probing field defined on a Fock space $\mathcal{F}_1$ . The total system $G_{T}$ can be described as
\begin{eqnarray}\label{47}
  G_{T}&=&({\rm I},\left(
                     \begin{array}{c}
                      \sqrt{\gamma_0}a_0 \\
                       {L} \\
                     \end{array}
                   \right)
,H_P+H_I+\omega_0a^\dagger_0a_0)
\end{eqnarray}
where $L$ is the coupling operator of the principal system for the probing field.

Note that supposing an operator of the augmented system can be denoted as $X'=X_p\otimes X_a$, the generator can be written as
\begin{eqnarray}
  \mathcal{G}_T(X') &=& \mathcal{G}_p(X_p)\otimes X_a+X_p\otimes  \mathcal{G}_a(X_a) \nonumber \\
&& -{\rm i}[X',H_I],
\end{eqnarray}
where
\begin{eqnarray}
  \mathcal{G}_p(X_p) &=&-{\rm i}[X_p,H_P]+\mathcal{L}_L(X_p), \\
  \mathcal{G}_a(X_a) &=& -{\rm i}[X_a,\omega_0 a_0^\dagger a_0]+\mathcal{L}_{\sqrt{\gamma_0}a_0}(X_a)
\end{eqnarray}
are the generators for the principal system and the ancillary system, respectively. The notation $\mathcal{L}_{\cdot}(\cdot)$ defines a Lindblad superoperator~\cite{xue2015}.

Here, the probing field satisfies a non-demolition condition and the detection efficiency is assumed to be $100\%$~\cite{bouten,belavkin}.
Hence, we can obtain a \textit{Belavkin} quantum filter for the augmented system as
\begin{eqnarray}\label{67}
    {\rm d}\pi_t( X')&=& \pi_t(\mathcal{G}_T (X')){\rm d}t-(\pi_t( X' L+{ L}^\dagger  X')-\pi_t( X')\nonumber\\
    &&\times\pi_t( L+ {L}^\dagger))({\rm d}Y_t-\pi_t( L+ {L}^\dagger)  {\rm d}t)
\end{eqnarray}
where the conditional expectation is denoted as $\pi_t(\cdot)$. The measurement results $Y(\tau),~0\leq\tau\leq t$ generate a commutative subspace $\mathcal{Y}_t$ and also drive an innovation process
${\rm d}W={\rm d}Y_t-\pi_t( L+ {L}^\dagger)  {\rm d}t$ which is equivalent to a classical Wiener process. Note that the increment ${\rm d}W$ is independent of $\pi_\tau( X'), 0\leq\tau\leq t$. Note that for linear principal systems, the \textit{Belavkin} filter is actually a quantum Kalman filter~\cite{PhysRevA.60.2700,PhysRevA.94.070405}.
\section{LQG feedback control of the Lorentzian-noise-disturbed linear non-Markovian quantum systems}\label{sec5}
\begin{figure}
  \includegraphics[width=8.5cm]{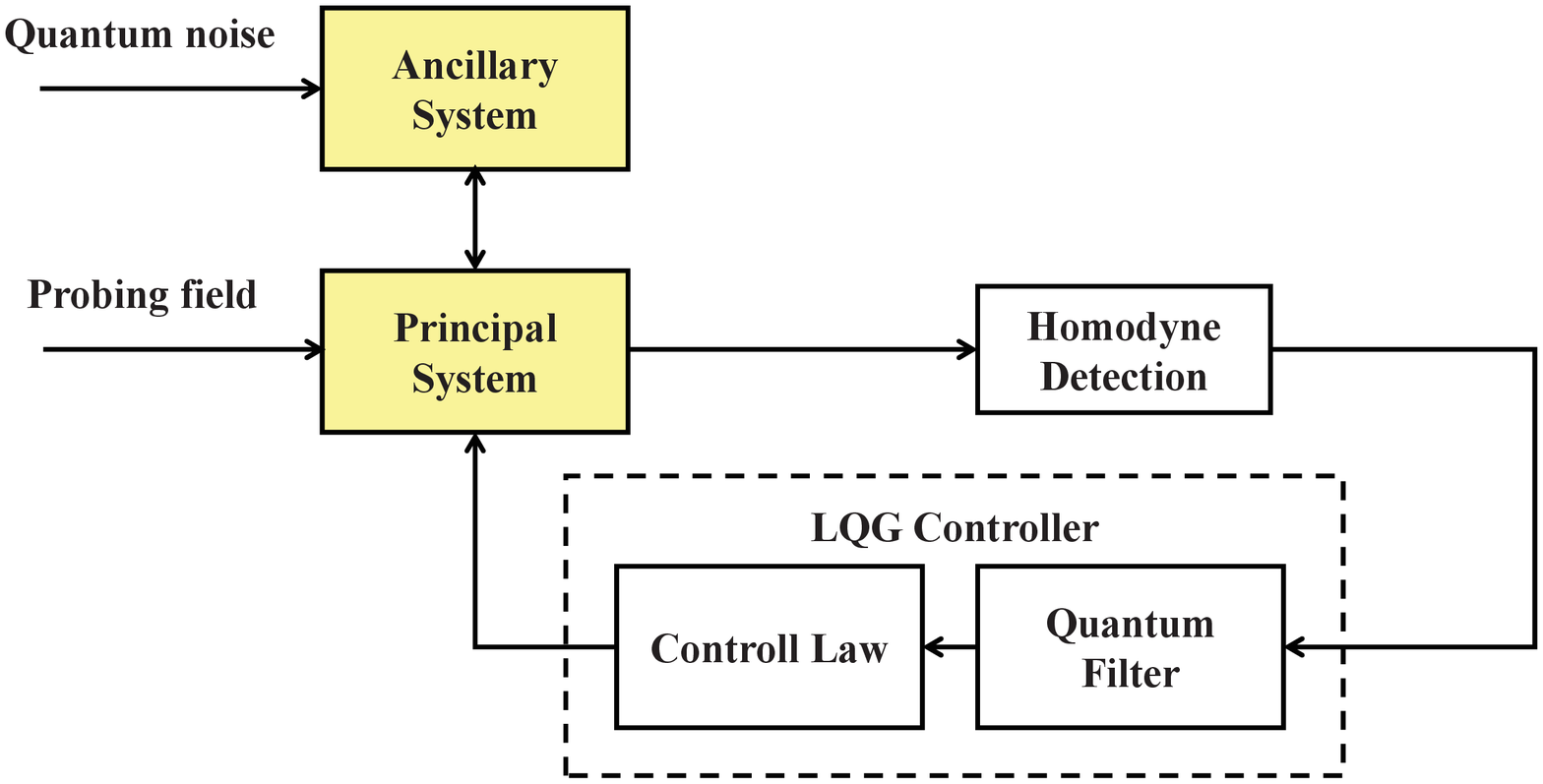}\\
  \caption{Block diagram of LQG feedback control of the non-Markovian quantum system.}\label{control}
\end{figure}
\subsection{The augmented system in a quadrature representation}


In this section, we may consider a more complicated case that the ancillary system is driven by thermal noise as well as quantum noise, since the noise spectrum generated by the ancillary system is independent on the input fields of the ancillary system. The thermal noise disturbs the ancillary system through the coupling operator  $\sqrt{\gamma_0}a_0$. Here, we denote the white quantum noise and thermal noise for the ancillary system as $b_w(t)$ and $b_n(t)$, respectively. In a stochastic description, the corresponding infinitesimal increment for white quantum noise ${\rm d}B$ satisfies ${\rm d}B{\rm d}B={\rm d}B^\dagger{\rm d}B={\rm d}B^\dagger{\rm d}B^\dagger=0$ and ${\rm d}B{\rm d}B^\dagger={\rm d}t$. However, the infinitesimal increment for thermal noise ${\rm d}A$ satisfies different relations as ${\rm d}A{\rm d}A={\rm d}A^\dagger{\rm d}A^\dagger=0$, ${\rm d}A{\rm d}A^\dagger=(1+N){\rm d}t$, and ${\rm d}A^\dagger{\rm d}A=N{\rm d}t$, where $N$ is the numbers of photons carried by the thermal noise~\cite{Gardiner, ryan}.

On the other hand, the controlled principal system we considered is an optical mode in a cavity, which is also a linear quantum system.
The Hamiltonian of the controlled cavity can be written as
\begin{equation}\label{}
 H_P=\omega_s a_s^\dagger a_s+\frac{u(t)}{2}(a_s^\dagger+a_s)
\end{equation}
with an angular frequency $\omega_s$ and an annihilation (creation) operator $a_s$ ($a_s^\dagger$), where $u(t)$ is a semiclassical control field. Then, the coupling operator $Z$ and $L$ can be specified as $Z=\sqrt{\kappa}a_s$ and $L=\sqrt{\gamma_1}a_s$.

Thus the Langevin equations for the augmented principal and ancillary cavities can be written as
\begin{eqnarray}\label{72-1}
&&\left[
        \begin{array}{c}
          \dot{a}_s(t)\\
         \dot{ a}_0(t)\\
        \end{array}
      \right]=\left[
                \begin{array}{cc}
                  -{\rm i}\omega_s-\frac{\gamma_1}{2} & \frac{\sqrt{\kappa\gamma_0}}{2} \\
                  -\frac{\sqrt{\kappa\gamma_0}}{2} & -{\rm i}\omega_0-\frac{\gamma_0}{2}\\
                \end{array}
              \right]\left[
        \begin{array}{c}
          a_s(t)\\
          a_0(t)\\
        \end{array}
      \right]\nonumber\\
      &&~~~~~-\left[
        \begin{array}{c}
          {\rm i}\frac{u(t)}{2}\\
          0\\
        \end{array}
      \right] -\left[
        \begin{array}{ccc}
          \sqrt{\gamma_1}&0&0\\
          0&\sqrt{\gamma_0}&\sqrt{\gamma_0}\\
        \end{array}
      \right]\left[
        \begin{array}{c}
          b_p(t)\\
          b_w(t)\\
          b_n(t)\\
        \end{array}
      \right],\nonumber\\
\end{eqnarray}
where $b_p(t)$ is the white noise probing field.
The output of the probing field $b_p(t)$ can be calculated as
\begin{equation}\label{72-2}
  b_{\rm out}(t)=b_p(t)+\sqrt{\gamma_1}a_s(t),
\end{equation}
which carries the information of the principal system.

It is convenient to move the Eqs.~(\ref{72-1}) and~(\ref{72-2}) to a quadrature representation with real-valued coefficient matrices as
\begin{eqnarray}
  \dot{ x}(t)&=& A { x}(t)+B u(t)+w_1(t)\label{72-2-1}\\
 y_p(t)&=&C{ x}(t)+w_2(t)\label{72-2-1-1}
\end{eqnarray}
by using a transformation matrix $\Xi=\frac{1}{\sqrt{2}}\left[
                          \begin{array}{cc}
                            1 & 1 \\
                            -{\rm i} &{\rm i} \\
                          \end{array}
                        \right]$.
Here, the coefficient matrices
\begin{eqnarray}
      A &=& \left[
              \begin{array}{cccc}
                -\frac{\gamma_1}{2} & \omega_s & \frac{\sqrt{\kappa\gamma_0}}{2} & 0 \\
                -\omega_s& -\frac{\gamma_1}{2} & 0 & \frac{\sqrt{\kappa\gamma_0}}{2} \\
                -\frac{\sqrt{\kappa\gamma_0}}{2}& 0 & -\frac{\gamma_0}{2} & \omega_0 \\
                0 & -\frac{\sqrt{\kappa\gamma_0}}{2} & -\omega_0 & -\frac{\gamma_0}{2} \\
              \end{array}
            \right],\\
            B&=&\left[
                    \begin{array}{cccc}
                      0& 1 & 0 & 0 \\
                    \end{array}
                  \right]^T,\\
    C&=& \left[
            \begin{array}{cccc}
             \sqrt{\gamma_1} &0 & 0 & 0 \\
            \end{array}
          \right],
    \end{eqnarray}
are real matrices and  the components of $ x(t)=[x_s^T(t),x_0^T(t)]^T=[ q_s(t),p_s(t), q_0(t),p_0(t)]^T$ are calculated as $x_s(t)=[ q_s(t),p_s(t)]^T=\Xi[a_s(t),a_s^\dagger(t)]^T$, $x_0(t)=[q_0(t),p_0(t)]^T=\Xi[a_0(t),a_0^\dagger(t)]^T$.
Since we observe the position component of the output field, the output is defined as $y_p(t)=\frac{1}{\sqrt{2}}(b_{\rm out}(t)+b_{\rm out}^\dagger(t))$.

In addition, the noise $w_1(t)$ and $w_2(t)$ are calculated as $w_1(t)=B'w(t)$ and $w_2(t)=D w(t)$, where the matrices $B'$ and $D$ are expressed as
\begin{eqnarray}
 &&B' =\left[
         \begin{array}{c|c}
           B'_1 & B'_2 \\
         \end{array}
       \right]=\nonumber\\
&&  \left[
          \begin{array}{cccc|cc}
           -\sqrt{\gamma_1} & 0 & 0 & 0 & 0 & 0 \\
            0 & -\sqrt{\gamma_1} & 0 & 0 &0 & 0 \\
            0 & 0 & -\sqrt{\gamma_0} & 0 & -\sqrt{\gamma_0} & 0 \\
            0 & 0 & 0 &-\sqrt{\gamma_0} & 0 & - \sqrt{\gamma_0} \\
          \end{array}
        \right], \nonumber\\
&&  D= \left[
         \begin{array}{c|c}
           D_1 & 0 \\
         \end{array}
       \right]=\left[
            \begin{array}{cccc|cc}
             1&0 & 0 & 0 &0&0\\
            \end{array}
          \right].
\end{eqnarray}
The components of $w(t)=[v_p(t),v_q(t),\nu_p(t),\nu_q(t),\mu_p(t),$ $\mu_q(t)]^T$ are calculated as $[v_p(t),v_q(t)]^T=\Xi[b_p(t),b_p^\dagger(t)]^T$, $[\nu_p(t),$ $\nu_q(t)]^T=\Xi[b_w(t),b_w^\dagger(t)]^T$, and $[\mu_p(t),$ $\mu_q(t)]^T=\Xi[b_n(t),b_n^\dagger(t)]^T$.

The white noise and thermal noise fields for the ancillary system and the probing field for the principal system are initially uncorrelated, leading to a covariance matrix of the noise $w(t)$ as
\begin{equation}\label{}
 M= \left[
    \begin{array}{cc}
      {\rm I}_{4\times4}& 0 \\
      0 & M_2\\
    \end{array}
  \right],
\end{equation}
where $M_2={\rm diag}[\frac{1}{2}+N,\frac{1}{2}+N]$ is the covariance matrix of the thermal noise and ${\rm I}_{4\times4}$ is a $4\times4$ identity matrix.

However, the noise $w_1(t)$ and $w_2(t)$ are correlated such that the intensity of the noise vector ${\rm col}[w_1(t),w_2(t)]$ can be calculated as
\begin{eqnarray}\label{intensity}
V&=&
\left[
             \begin{array}{cc}
               V_1& V_{12} \\
               V_{12}^T & V_2 \\
             \end{array}
           \right]\nonumber\\
           &=& \left[
            \begin{array}{cc}
              B'_1{B}^{\prime T}_1+B'_2M_2B_2^{\prime T}& B'_1D_1^T\\
              D_1{B}_1^{\prime T}&D_1D_1^T\\
            \end{array}
          \right].
\end{eqnarray}
Note that there are non-zero elements in the off-diagonal block $V_{12}$ and $V_2$ is invertible.

\subsection{The whitening quantum Kalman filter}
For the linear model of the principal and ancillary system~(\ref{72-2-1}) and~(\ref{72-2-1-1}), the quantum filter (\ref{67}) is a quantum Kalman filter~\cite{PhysRevA.60.2700,PhysRevA.94.070405} and can be expressed as

\begin{eqnarray}\label{kalmanfilter}
  \dot{\hat x}_t &=& A\hat x_t  +Bu(t)+K(y_p-C\hat x_t), \nonumber\\
  0&=& (A-V_{12}V_2^{-1}C)\hat V_t+\hat V_t(A-V_{12}V_2^{-1}C)^T-\nonumber\\
  &&\hat V_tC^TV_2^{-1}C\hat V_t+V_1-V_{12}V_2^{-1}V_{12}^T,
\end{eqnarray}
where $\hat x_t$ is the estimate of the state vector $x(t)$ and the Kalman gain is $K=(\hat V_tC^T+V_{12})V_2^{-1}$. The conditional dynamics for $\hat x_t$ are driven by the error covariance matrix ${\hat V}_t$. Also, the noise in the state equation~(\ref{72-2-1}) and that in the output equaiton~(\ref{72-2-1-1}) are correlated, i.e., $V_{12}$ is non-zero.
\subsection{LQG control problem}
Based on the estimation of the states of both the principal and ancillary systems, a semiclassical feedback control law $u(t)$ can be designed by solving a LQG control problem. The LQG problem for this non-Markovian quantum system can be formulated as follows:

Consider the Markovian representation of a non-Markovian quantum system (\ref{72-2-1}) whose output (\ref{72-2-1-1}) can be used to construct the quantum Kalman filter~(\ref{kalmanfilter}). The joint white noise process ${\rm col}[w_1(t),w_2(t)]$ has the intensity (\ref{intensity}). The LQG feedback control problem is  to find  the feedback control
$u(t)$ based on the estimate $\hat x(t)$, such that the objective
\begin{eqnarray}\label{obj}
 &&J(u(t))=\lim_{T\rightarrow \infty}\frac{1}{T}{\Big \langle}\int_0^T(x_s^T(t)Q_1x_s(t)+Q_2u^2(t)){\rm d}t+\nonumber\\
 &&~~~~~~~~~~~~~~~~~~~~~~~~~~~~~~x_s^T(T)Q_3x_s(T){\Big \rangle}
\end{eqnarray}
is minimized. Here, $Q_1$, $Q_2$, and $Q_3$ are symmetric weighting matrices such that $Q_1>0$, $Q_2>0$, and $Q_3>0$ and the final time is denoted as $T$. Note that the objective $J$ is a function of the states of the principal system $x_s(t)=Ex(t)$ except the states of the ancillary system $x_0(t)$. The ancillary system plays the role of the internal modes of the environment such that the states $x_0(t)$ are uncontrollable.
\subsection{LQG feedback control}
The optimal feedback control law $u^*(t)$ to minimize the objective $J$ over the times can be obtained by using the dynamical programming method~\cite{KS72,ryan}. It turns out that the optimal feedback control can be expressed as
\begin{equation}\label{controlu}
  u^*(t)=-F\hat x(t)
\end{equation}
with $F=Q_2^{-1}B^TP_\infty$. The control $u^*(t)$ is generated based on the estimated state $\hat x(t)$ of the augmented system. The symmetric matrix $P_\infty$ is a solution of the Riccati equation
%
\begin{equation}\label{}
0=P_\infty A+A^TP_\infty-P_\infty BQ_2^{-1}B^TP_\infty+E^TQ_1E.
\end{equation}
Note that the separation principle implies that the controller and the filter can be designed separately~\cite{KS72}.
\subsection{Closed loop dynamics}
By applying the feedback control to the principal system, we can obtain the dynamics of the closed loop system including the controlled plant and the controller. Denoting the closed loop state as $\tilde{x}(t)=\left[
                                                \begin{array}{cc}
                                                  x^T(t)& \hat x^T(t) \\
                                                \end{array}
                                              \right]^T
$, the closed loop system dynamical equation can be written as
\begin{equation}
 \dot{\tilde{x}}(t) = \tilde{A}\tilde{x}+\tilde{B}w(t)
 \end{equation}
 where
\begin{eqnarray}
\tilde{A} &=& \left[
                   \begin{array}{cc}
                     A & -BF \\
                     KC & A-KC-BF \\
                   \end{array}
                 \right],\\
 \tilde{B} &=& \left[
                 \begin{array}{c}
                   B' \\
                   KD \\
                 \end{array}
               \right].
\end{eqnarray}
And the covariance matrix $\tilde{P}$ of the closed loop system in a steady state satisfies a Lyapunov equation
\begin{equation}\label{ly}
   \tilde{A} \tilde{P}+ \tilde{P}\tilde{A}^T+\tilde{B}M\tilde{B}^T=0.
\end{equation}

\section{Example}\label{sec6}
\begin{figure}
  \includegraphics[width=8.5cm]{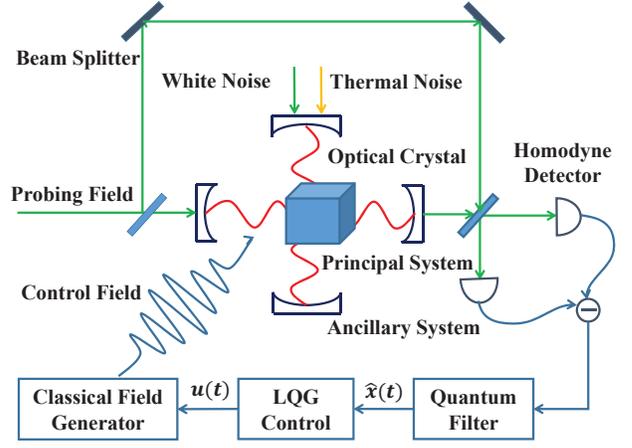}\\
  \caption{Possible experimental realization of the LQG controlled non-Markovian quantum system.}\label{example}
\end{figure}
A possible experimental realization of the controlled non-Markovian quantum system is shown in Fig.~\ref{example}. The principal system is an optical mode in a cavity, which is directly coupled with an optical mode in an ancillary cavity via an optical crystal. The ancillary system is driven by both white quantum noise and thermal noise. A probing field is injected to the principle system, whose output is observed via Homodyne detection to determine a whitening filter. The estimated state is utilized by the LQG control law.

Since the ancillary system is driven by thermal noise, the principal system will be also heated due to the direct interaction; i.e., the photon numbers in the principal system will increase. Hence, we apply LQG feedback control to cool the principal system, for which the objective can be written as
\begin{equation}\label{obj1}
  J=\langle a_s^\dagger a_s\rangle=\langle x^T(t)E^TQ_1Ex(t)\rangle
\end{equation}
where $Q_1={\rm diag}[\frac{1}{2},\frac{1}{2}]$ and $E=[{\rm I}_{2\times2},0]$ with a $2\times2$ identity matrix ${\rm I}_{2\times2}$.
Alternatively, the objective (\ref{obj1}) in the steady state can be calculated as
\begin{equation}\label{}
  J={\rm tr}[\tilde{P}\tilde{E}^TQ_1\tilde{E}]
\end{equation}
with $\tilde{E}=\left[
                  \begin{array}{cccc}
                    {\rm I}_{2\times2} & 0 & 0 & 0 \\
                  \end{array}
                \right],
$
where $\tilde{P}$ is the solution of the Lyapunov equation (\ref{ly}) for the closed loop system.

\begin{figure}
  \includegraphics[width=8.5cm]{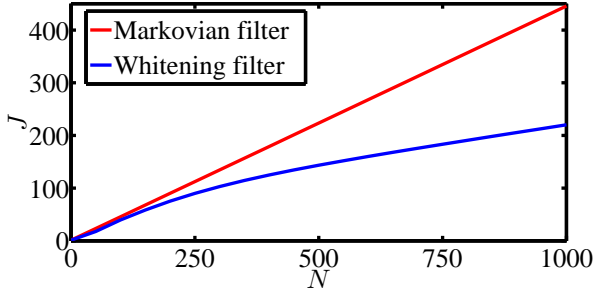}\\
  \caption{The photon numbers of the principal system $J$ as a function of $N$, where the red and blue lines represent the LQG controller with a Markovian filter and a whitening filter, respectively.}\label{J}
\end{figure}

Fig. \ref{J} shows the steady state photon numbers of the principal system; i.e., the objective $J$ varies with the photon number of the thermal noise $N$. In this simulation, the angular frequencies of the principal and ancillary system are assumed to be $\omega_s=\omega_0=10{\rm GHz}$. The damping rates of the ancillary system with respect to the noise and that of the principal system with respect to the probing field are $\gamma_0=\gamma_1=1$. The coupling strength between the principal and ancillary system is $\kappa=2$. In addition, the weighting matrix $Q_2$ is assumed to be far smaller than the element of $Q_1$, e.g., $Q_2=0.05$.
The simulation results show that the LQG controller with a whitening filter has a better performance than that with a Markovian filter. As the strength of the thermal noise increases, the objective for the controller with a Markovian filter increases linearly. The derivation of the LQG controller with a Markovian filter is given in the Appendix. On the contrary, the controller with the whitening filter can achieve better performance. Since the whitening filter can estimate the state of both the principal and ancillary systems, the controller with the whitening filter can give the optimal control field. However, the controller with the Markovian filter lacks information on the ancillary system. As a result, its control performance degrades when the effect of the ancillary system becomes large.

\section{Conclusion}\label{sec7}

In this paper, we have presented a LQG feedback strategy to control a Lorentzian-noise-disturbed linear non-Markovian quantum system based on its Markovian representation model. In this model, the ancillary systems play the role of the internal modes of the non-Markovian environment converting white noise to Lorentzian noise and thus the principal system obeys a non-Markovian dynamics due to its direct interaction with the ancillary system. With this Markovian representation model,  the states of the augmented system can be continuously estimated by a whitening filter. The LQG feedback control is designed for the non-Markovian system based on the results of the whitening filter. Simulation results show the control performance is enhanced when we substitute the Markovian filter by the whitening filter. For future work, it is possible to apply this feedback control strategy to more general non-Markovian quantum systems with arbitrary noise spectra.

\appendix
\subsection{System model for designing a Markovian quantum filter}
If the whitening filter is substituted by a Markovian quantum filter, the control performance may degrade.
Supposing we design the quantum filter based on a Markovian system model only; i.e., we only consider the probed principal system, the dynamical equations in the quadrature representation can be written as
\begin{eqnarray}
  \dot{ x}_m(t)&=& \bar{A} { x_m}(t)+\bar{B} u(t)+\bar{w}_1(t)\\
 y_p(t)&=&\bar{C}{ x_m}(t)+\bar{w}_2(t),
\end{eqnarray}
with
\begin{eqnarray}
    && \bar A =\left[
              \begin{array}{cc}
                -\frac{\gamma_1}{2} & \omega_p \\
                -\omega_p& -\frac{\gamma_1}{2}
              \end{array}
            \right],~~~~
     \bar B =\left[
               \begin{array}{c}
                 0  \\
                 1
               \end{array}
             \right]
     , \nonumber\\
    &&  \bar  C=[\sqrt{\gamma_1},0],~~~~~~~~~~~~
\end{eqnarray}
where ${ x_m}(t)=\left[
                   \begin{array}{cc}
                     q_m(t) & p_m(t) \\
                   \end{array}
                 \right]^T
$ is the state of the principal system. The angular frequency is denoted as $\omega_p$. In the simulation, we let $\omega_p=10{\rm GHz}$.
The noise terms can be further expressed as $\bar{w}_1(t)=\bar B'\bar{w}(t)$ and $\bar{w}_2(t)=\bar D\bar{w}(t)$ with
$$
\bar B'={\rm diag}[-\sqrt{\gamma_1},-\sqrt{\gamma_1}],~~~~\bar D=[1,0],\nonumber
$$
where $\bar{w}(t)=[v_p(t),v_q(t)]^T$ is the noise of the probing field. Note that the noise $\bar w_1(t)$ and $\bar w_2(t)$ are correlated and their covariance matrix can be calculated as
\begin{eqnarray}\label{cov2}
  &&\mathbb{E}{\Big [}\left[
               \begin{array}{c}
                 \bar w_1(t) \\
                 \bar w_2(t) \\
               \end{array}
             \right]\left[
                      \begin{array}{cc}
                       \bar w_1^T(t) &\bar w_2^T(t)\\
                      \end{array}
                    \right]
  {\Big ]}\nonumber\\
  &=&\left[
             \begin{array}{cc}
               \bar V_1& \bar V_{12} \\
               \bar V_{12}^T & \bar V_2 \\
             \end{array}
           \right]= \left[
            \begin{array}{cc}
              \bar B'{{\bar B}}^{\prime T}& \bar B'\bar D^T\\
              \bar D{{\bar B}}^{\prime T}&\bar D\bar D^T\\
            \end{array}
          \right].
\end{eqnarray}
Here, $\bar V_2$ is invertible and $\bar V_{12}\neq 0$.
\subsection{Markovian quantum filter}
Based on this Markovian model, we can obtain a Markovian filter as
\begin{eqnarray}
  \dot{\hat x}_m(t) &=& \bar{A}\hat x_m(t)  +\bar{B}u(t)+\bar{K}(y_p-\bar{C}\hat x_m(t)), \nonumber\\
  0&=& (\bar{A}-\bar{V}_{12}\bar{V}_2^{-1}\bar{C})\hat V_m(t)+\nonumber\\
  &&\hat V_m(t)(\bar{A}-\bar{V}_{12}\bar{V}_2^{-1}\bar{C})^T+\bar{V}_1-\nonumber\\
  &&\hat V_m(t)\bar{C}^T\bar{V}_2^{-1}\bar{C}\hat V_m(t)-\bar{V}_{12}\bar{V}_2^{-1}\bar{V}_{12}^T,
\end{eqnarray}
where the estimated state is denoted as $\hat x_m(t)$ and its covariance matrix as $\hat V_m(t)$. The gain of the filter can be calculated as $\bar{K}=(\hat V_m(t)\bar{C}^T+\bar{V}_{12})\bar{V}_2^{-1}$.
\subsection{LQG control based on the Markovian filter}
The LQG feedback control law can be designed as
\begin{equation}\label{uu}
 u(t)=-\bar{F}\hat x_m(t),
\end{equation}
by using the estimated state of the Markovian filter $\hat x_m(t)$, where $\bar F=Q_2^{-1}\bar B^T\bar P_\infty$. $\bar P_\infty$ is a solution of the Riccati equation
\begin{equation}\label{riccati2}
0=\bar P_\infty\bar A+\bar A^T\bar P_\infty-\bar P_\infty\bar BQ_2^{-1}\bar B^T\bar P_\infty+Q_1.
\end{equation}
\subsection{Closed loop system dynamics}
By applying the control (\ref{uu}) to the non-Markovian system model (\ref{72-2-1}) and (\ref{72-2-1-1}), the closed loop system dynamical equation can be obtained as
\begin{equation}
 \dot{\tilde{x}}_m(t) = \tilde{A}'\tilde{x}_m+\tilde{B}'w(t),
 \end{equation}
 where $\tilde{x}_m(t)=\left[
                                                \begin{array}{cc}
                                                  x(t)& \hat x_m(t) \\
                                                \end{array}
                                              \right]^T
$ and the coefficient matrices are written as
 \begin{eqnarray}
  \tilde{A}' &=& \left[
                   \begin{array}{cc}
                     A & -B\bar{F} \\
                     \bar{K}C & \bar{A}-\bar{K}\bar{C}-\bar{B}\bar{F} \\
                   \end{array}
                 \right],
   \\
 \tilde{B}' &=& \left[
                 \begin{array}{c}
                   B' \\
                   \bar{K}D \\
                 \end{array}
               \right].
\end{eqnarray}

The covariance matrix $\tilde{P}_m$ for the steady state of the closed loop system satisfies a Lyapunov equation as
\begin{equation}\label{}
   \tilde{A}' \tilde{P}_m+ \tilde{P}_m{\tilde{A}}^{'T}+\tilde{B}_mM{\tilde{B}}_m^T=0.
\end{equation}
And thus the objective in a steady state can be calculated as
\begin{equation}\label{}
  J={\rm tr}[\tilde{P}_m\tilde{E}^{'T}Q_1\tilde{E}']
\end{equation}
with $\tilde{E}'=\left[
                  \begin{array}{ccc}
                    I_{2\times2} & 0 & 0 \\
                  \end{array}
                \right].
$

\end{document}